\documentclass[letterpaper,10pt]{article} 

\usepackage{opticameet3} 

\newcommand\authormark[1]{\textsuperscript{#1}}

\usepackage{amsmath,amssymb}
\usepackage[colorlinks=true,bookmarks=false,citecolor=blue,urlcolor=blue]{hyperref} 

\begin{document}

\title{New Interferometric Testing Utility (NITU) : A Python Package for Interferometric Data Analysis and Visualization }


\author{Meghdoot Biswas\authormark{1,*}and Daewook Kim\authormark{1,2}}

\address{\authormark{1} James C. Wyant College of Optical Sciences, University of Arizona, 1630 E. University Blvd.,
Tucson, AZ 85721 USA}

\address{\authormark{2}Department of Astronomy, University of Arizona, 933 N. Cherry Ave., Tucson, AZ, 85721 USA}

\email{\authormark{*}meghdoot@arizona.edu} 

\begin{abstract}
New Interferometric Testing Utility (NITU) is a newly developed Python package for analyzing and visualizing interferometric data. It provides Zernike decomposition, interactive visualization, time series analysis, and additional features for optical manufacturing and testing.
\end{abstract}

\section{Introduction}

In this report, we introduce New Interferometric Testing Utility (NITU), a newly developed Python package specifically designed for interferometric data analysis and visualization. NITU is “new” in terms of its features, functionality and also the platform it uses. While Poppy\cite{perrin2016poppy} includes some Zernike functionalities, NITU offers a broader array of specialized tools for interferometric data analysis and visualization, enhancing its capabilities beyond those provided by Poppy. A MATLAB based open source package Saguaro\cite{kim2011open} offers some great features to analyze and visualize interferometric data but being a MATLAB based package, it has some demerits. A notable one is, one has to have a MATLAB license in order to use it. On the other hand, NITU being a Python based package, is free. Python’s better readability leads to fewer bugs and faster debugging. Also Python offers a wider set of choices in graphics packages and toolsets which would make it available for a greater community\cite{ozgur2017matlab}. We tried to make NITU with minimal external dependencies for improved performance and easier maintenance. To this date, it only uses NumPy, SciPy, Matplotlib and Plotly\cite{plotly} to provide an interactive environment.  NITU offers some unique features such as time series analysis of interferometric data which is crucial in optical manufacturing/testing where the unit under test is needed to be monitored during the process to achieve highest quality. Also to measure the temporal variability of the surface that can change by several factors such as temperature, vibration to name a few.
We highlight several key features of NITU that are rigorously utilized by optical metrology experts in both academic and industrial settings. We also discuss the advantages of Python over MATLAB for NITU and outline the future development plans for the package.

\section{Package Highlights}
\subsection{Zernike decomposition}
One of the most important and widely utilized tools in interferometry is Zernike decomposition and Zernike fitting. NITU employs Noll’s indexing for Zernike decomposition\cite{noll1976zernike}. Figure \ref{fig:nitu} (a) shows the input optical path difference (OPD) map and the reconstructed wavefront using least squares fitting for the first 37 Zernike polynomials.

\subsection{Piston Tip and Tilt removal}
Removal of piston, tip, and tilt from the OPD is often required in optical metrology to cancel out the aberrations induced by misalignment which are not the surface property of the unit under test. The \texttt{remove\_ptt()} function of NITU can remove this piston, tip and tilt from the reconstructed OPD map. Figure \ref{fig:nitu} (b) demonstrates the implementation of the function

\subsection{Time series analysis}

NITU offers a \texttt{timeseries()} function that fits the root mean square (RMS) of the surface or a specific Zernike polynomial change with a straight line. It returns the equation of the fitted line and the standard deviation of the data.

\subsection{Interactive visualization}
\texttt{InteractiveMode()} allows users to view, zoom, and rotate data in three dimensions simply by moving the cursor, without requiring additional lines of code. This interactive capability is particularly useful for identifying anomalies and gaining a clearer understanding of the data. Users can easily customize their view to meet specific needs, enhancing the overall data analysis experience.
The function's implementation is illustrated in Figure \ref{fig:nitu} (c).

\subsection{The "summary"}
\begin{figure} [ht]
   \begin{center}
   \begin{tabular}{c} 
   \includegraphics[height=5cm]{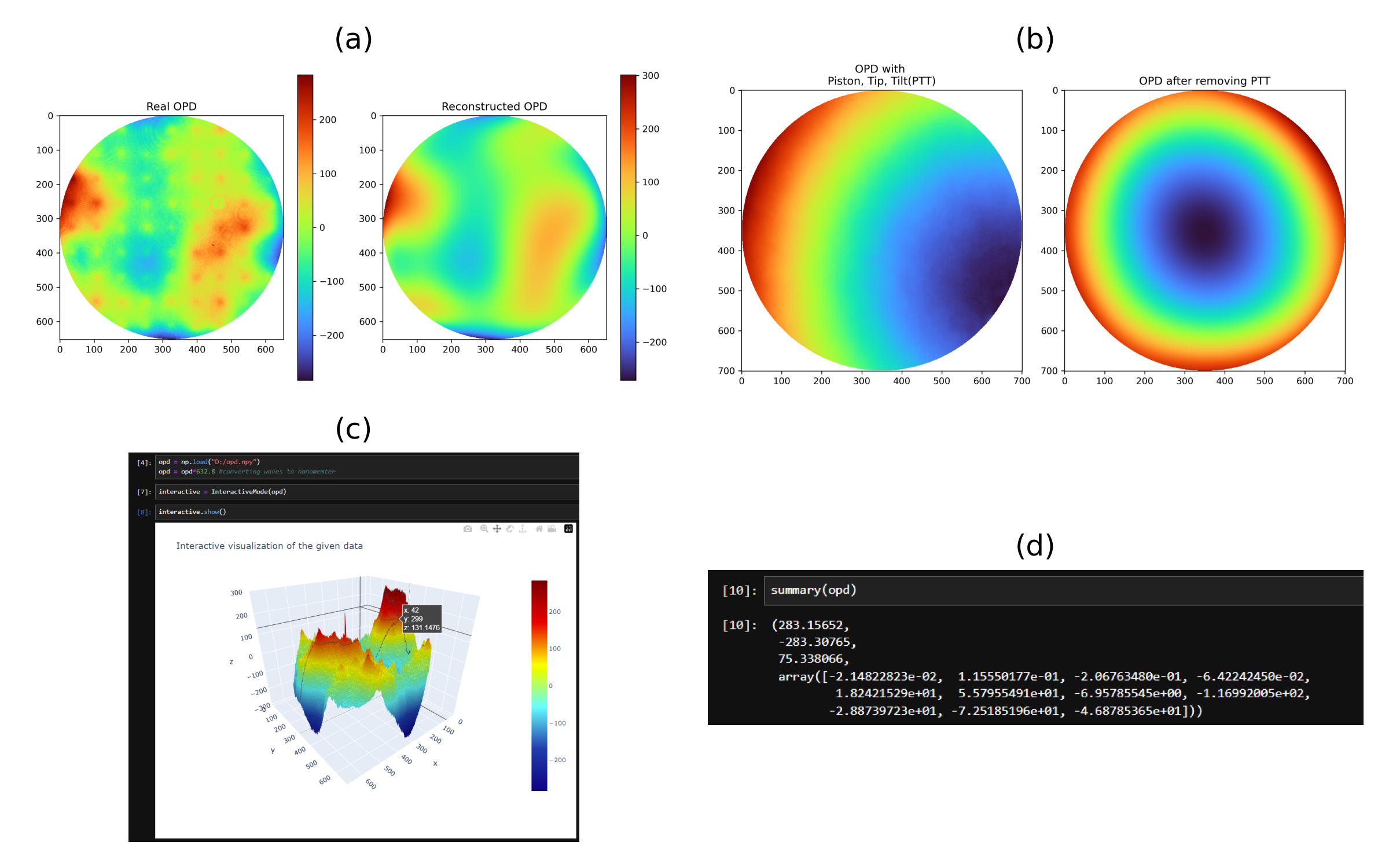}
   \end{tabular}
   \end{center}
   \caption[example] 
   { \label{fig:nitu} 
(a) Comparison of the input OPD with the reconstructed OPD. (b) OPD after removing Piston, Tip, and Tilt. (c) Interactive visualization of the OPD data. (d) Implementation details of the summary function.}
    \end{figure}

The \texttt{summary()}, function offers a quick overview of a given OPD map. It provides a brief summary of the input data, including the peak and valley values, the surface RMS, and the first 11 Zernike coefficients. This function is essential for obtaining key metrics and insights about the OPD efficiently. Figure \ref{fig:nitu} (d) depicts how the function is implemented.

\section{Key Reasons to Prefer Python Over MATLAB}

Choosing Python over MATLAB has several advantages. Python is free and open-source, which eliminates cost barriers and makes it accessible to everyone. Its straightforward syntax enhances code readability and reduces bugs. Unlike MATLAB, which is designed primarily for matrix manipulation, Python is versatile and supports a wide range of libraries, lists, and dictionaries for efficient coding. Python’s zero-based indexing, in line with most programming languages, reduces confusion compared to MATLAB’s one-based indexing. Python also excels in object-oriented programming(OOP) with a clean, flexible structure, whereas MATLAB’s OOP can be more complex. Additionally, Python offers a diverse range of graphical packages for creating appealing and functional applications. These features make Python a superior choice for academic and industrial applications in optical metrology and beyond.\\
This section is based on the findings of Ozgur, Ceyhun, et al.\cite{ozgur2017matlab}

\section{Future Work}

We plan to make NITU an open-source Python package, allowing the scientific and engineering community to access and contribute to its development. This will enable collaboration from researchers and developers worldwide, leading to new features and improvements.


\bibliography{sample}
\bibliographystyle{opticajnl}
\end{document}